# Image polaritons in van der Waals crystals

Sergey G. Menabde[1], Jacob T. Heiden[1], Joel Cox[2,3], N. Asger Mortensen[2,3,4,*], and Min Seok Jang[1,**]

[1] School of Electrical Engineering, Korea Advanced Institute of Science and Technology, Daejeon 34141, Korea
[2] Center for Nano Optics, University of Southern Denmark, DK-5230 Odense, Denmark
[3] Danish Institute for Advanced Study, University of Southern Denmark, DK-5230 Odense, Denmark
[4] Center for Nanostructured Graphene, Technical University of Denmark, DK-2800 Kongens Lyngby, Denmark

[*] asger@mailaps.org
[**] jang.minseok@kaist.ac.kr

**Abstract:**

Polaritonic modes in low-dimensional materials enable strong light-matter interactions and provide a platform for light manipulation at nanoscale. Very recently, a new class of polaritons has attracted considerable interest in nanophotonics: image polaritons in van der Waals crystals, manifesting when a polaritonic material is in close proximity to a highly conductive metal, so that the polaritonic mode couples with its mirror image. Image modes constitute an appealing nanophotonic platform, providing an unparalleled degree of optical field compression into nanometric volumes while exhibiting lower normalized propagation loss compared to conventional polariton modes in van der Waals crystals on non-metallic substrates. Moreover, the ultra-compressed image modes provide access to the nonlocal regime of light-matter interaction. In this Review, we systematically overview the young yet rapidly growing field of image polaritons. We discuss their dispersion properties, showcase the diversity of image modes in various van der Waals materials, and highlight the experimental breakthroughs owing to the unique properties of image polaritons.

## 1. Introduction

Low-dimensional van der Waals (vdW) crystals support a variety of polaritons – hybrid quasiparticles stemming from the coupling of light to the dipole moment associated with collective oscillations in matter [1-3]. A single atomic layer of a vdW crystal represents the limiting case of a free-standing polaritonic waveguide, where propagating surface modes obtain the shortest wavelength in a given material. The most comprehensively studied case of such 2D polaritons is graphene surface plasmons (GSPs) [4-6]. Surface plasmons are also predicted to be found in monolayer black phosphorus [7-9], while 2D phonon-polaritons have been observed in monolayer hexagonal boron nitride (hBN) [10, 11]. At the same time, multilayer vdW crystals are inherently anisotropic, and support hyperbolic phonon-polaritons (HPPs) within the spectral bands for which the real part of the permittivity tensor has components with opposite sign [2, 12, 13]. Such modes have been widely studied in hBN slabs [14-26], and have recently been discovered in orthorhombic molybdenum trioxide ($\alpha$-MoO$_3$) [27, 28].

Shortly following the discovery of graphene, a theoretical investigation of the electrodynamic response of two spatially separated graphene layers revealed that such a double-layer system supports two modes with opposite field symmetry in the out-of-plane electric field component $E_z$ [29], as schematically illustrated in the middle panel of Fig. 1A. The tightly confined symmetric mode was termed an "acoustic" graphene plasmon (AGP) due to its linear dispersion at high frequencies. Notably, the AGP mode does not experience a cutoff when the separation between layers approaches zero, resembling the plasmonic mode of a narrow metal slit. Importantly, the symmetric field distribution of the AGP allows it to manifest in graphene separated by a thin dielectric spacer from a highly conductive metal, with the metal behaving as a perfect electric conductor (PEC) at sufficiently low optical frequencies [30, 31]. In such a heterostructure, the image charges in the metal effectively "reflect" the collective oscillations of electrons in graphene, resulting in an image graphene plasmon (IGP) mode equivalent to the AGP, where the metal surface is the symmetry plane (Fig. 1A). Because the IGP does not have a geometry-associated cutoff, it can be confined within a very thin (< 1 nm) dielectric spacer, where light is compressed on length scales down to ~$10^5$ times below its associated free-space wavelength [32], thus granting opportunities to probe phenomena beyond classical electrodynamics, including nonlocal phenomena in both the graphene and the metal [33-36]; the IGP also holds the potential to explore collective modes in strongly-correlated electron systems [37]. When compared to GSPs in the classical limit, IGPs exhibit a reduced sensitivity to intrinsic losses in graphene, which is a consequence of the IGP mode being not strictly bound to the graphene layer, so that its energy instead flows mainly within the low-loss dielectric spacer [38].

In the case of HPPs in a multilayer vdW crystal (i.e., a polaritonic slab waveguide), the dispersion solution contains an infinite number of modes with alternating field symmetry [14, 39]. While the first-order mode is anti-symmetric with respect to $E_z$, the second-order mode is symmetric, as shown in Fig. 1B. Then, in the specific case of hBN on a PEC-like metal, the symmetric second-order mode in the slab is equivalent to the fundamental mode (Fig. 1B). Here, as in the case of the IGP, the polariton mode in the slab couples to its image in the metal, forming a hyperbolic image phonon-polariton (HIP) mode. Thus far, the HIP has been experimentally observed only in hBN, while it is also anticipated to appear in $α$-MoO$_3$.

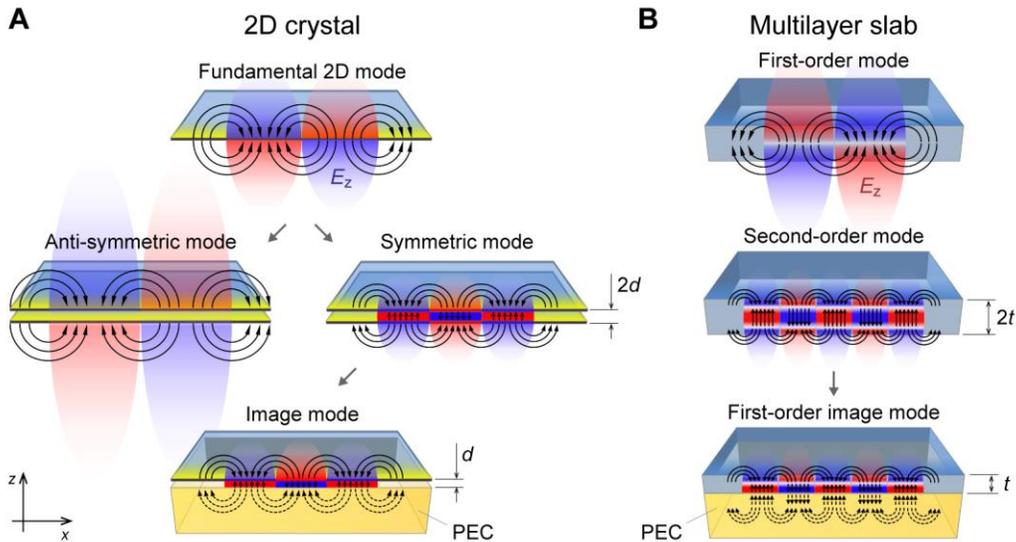

**Figure 1:** Two types of image polaritons in van der Waals crystals. Black curves schematically show the instantaneous electric field of the propagating polariton mode, and colors depict the evanescent $z$-component of the field, which can be detected by near-field microscopy. **A** Image mode in a 2D sheet, corresponding to the symmetric mode in a system of two closely placed monolayers of a van der Waals material separated by a dielectric spacer of thickness $d$. **B** Image mode in a multilayer slab, corresponding to the second-order waveguide mode in a van der Waals crystal that has twice the thickness $t$ of its counterpart interfacing a perfect electric conductor (PEC).

The image modes of plasmon- and phonon-polaritons have been predicted to exist in various vdW materials, and do not necessarily exhibit a linear dispersion. Furthermore, use of the term "acoustic" in the context of phonon-polaritons may be misleading. Therefore, in this Review, we consistently use the universal term "image polaritons" to describe the corresponding modes in different materials, as also suggested by Lee *et al.* [40].

Throughout the Review, we emphasize that image polaritons exhibit much stronger field compression and longer propagation length in optical cycles (i.e., lower normalized propagation loss) compared to their counterparts in vdW crystals on non-metallic substrates. Due to these overarching advantages across multiple materials and different polaritonic species, image polaritons present an appealing new platform in which to simultaneously harness exceptionally strong light-matter interaction and wave phenomena for nanophotonics applications, while further providing access to the nonlocal optical response regime.

The Review is organized as follows: first, in Section 2, we discuss theoretical aspects of polariton dispersion in vdW crystals, both for 2D monolayer and multilayer slab geometries. Then, in Section 3, we provide an overview of experimental studies on mid-infrared (mid-IR) IGPs, including near- and far-field probing, cases of extreme optical field compression, and an example of IGPs applied in molecular sensing. Section 4 highlights near- and far-field investigations of mid-IR HIPs in hBN, while Section 5 surveys state-of-the-art theoretical and experimental investigations of nonlocal electrodynamics in graphene and metals that becomes accessible via the ultra-compressed IGP. In Section 6 we turn to exciton-polaritons in semiconducting vdW materials interfacing metal substrates, focusing on the hybridization of excitons with surface plasmons. Finally, in Section 7, we provide an outlook for the emerging field of image polaritons.

## 2. Dispersion of image polaritons

We start with a discussion on the dispersion properties of image polaritons, presenting an analytical approximation of the dispersion relations derived for mono- and multilayer vdW crystals. So far, image plasmons in graphene and black phosphorus, along with image phonon-polaritons in hBN and $\alpha$-MoO$_3$, have been studied theoretically. In the classical regime, all image modes are predicted to have both significantly larger momentum and longer normalized propagation length, independently of their species.

**2.1 Image modes in 2D crystals**

The dispersion relation that characterizes image polaritons in a monolayer vdW material can be obtained by searching for the eigenmodes of a layered system comprised of a metal plate, a dielectric spacer, and a 2D layer which satisfy the electromagnetic boundary conditions [41]. Within classical electrodynamics, the dispersion can be derived by treating the vdW crystal as a 2D sheet with optical conductivity $\sigma(\omega)$ and the metal as a PEC to yield the closed-form dispersion of image polaritons [42, 43]:

$$\mathrm{i}\tan(k_{2,z}d) = \frac{k_{1,z}}{k_{2,z}}\frac{\varepsilon_2}{\varepsilon_1 + 4\pi k_{1,z}\sigma/\omega},$$

where $k_{j,z} = \sqrt{\varepsilon_j k_0^2 - q^2}$ is the $z$-component of the polariton wavevector in medium $j$ with relative permittivity $\varepsilon_j$, $q$ is the polariton propagation constant, and $k_0 = \omega/c$ is the free-space wave vector; here the index $j = 1$ corresponds to the half-space above the 2D layer, while $j = 2$ corresponds to the dielectric spacer of thickness *d*. Under the approximation of a thin spacer, such that $\text{Re}\{q\}d \ll 1$, and large momentum $q \gg k_0$ (i.e., the quasistatic limit), the dispersion in vacuum ($\varepsilon_1 = \varepsilon_2 = 1$) takes a simple form, which is typically a good approximation for highly confined image modes [42, 43]:

$$q = \sqrt{\frac{i\omega}{4\pi d\sigma}}. \quad (1)$$

The dispersion relation in Eq. (1) provides a great deal of insight into the main properties of image polaritons. First, a vanishingly thin spacer leads to a diverging polariton wavenumber and field amplitude inside the spacer, without any cutoff, such that the compression of polaritonic modes into an atomically thin dielectric layer becomes possible. Second, the normalized propagation length of image polaritons, associated with the in-plane wavevector *q*, increases as *d* decreases. The propagation length is traditionally used as a figure of merit (FOM) for propagating polaritons, and is defined in terms of optical cycles as $L/\lambda_p = \text{Re}\{q\}/2\pi\text{Im}\{q\}$, where *L* denotes the absolute propagation length and $\lambda_p$ the polariton wavelength. In the opposite limit of an infinitely thick spacer, the dispersion of image modes merges with that of the surface modes in a single-layer [42]. We emphasize that the divergence in wave vector associated with vanishing *d* is a consequence of classical electrodynamics in the quasistatic approximation and is actually compensated by a possible nonlocal response, as we will revisit in Section 5.

Thus far, the most comprehensively studied image mode has been the IGP, which has been observed in both far- and near-field experiments that will be discussed in Section 3. The numerically calculated field profile of the IGP is shown in Fig. 2A, and its analytically and numerically calculated dispersion and FOM at mid-IR frequencies are shown in Fig. 2B, as originally calculated by Voronin *et al.* [42]. Both the momentum and the FOM strongly depend on the spacer thickness, and the larger FOM at higher frequencies can be explained by a more rapid reduction of the IGP wavelength compared to its group velocity.

Besides the inherently larger FOM stemming from the dispersion, the field of a compressed IGP mode in a thin dielectric spacer is mainly localized within the spacer, in contrast to GSPs, for which the field is maximal at the graphene plane (Fig. 2A). The field localization of the IGP thus leads to a different spatial distribution of the mode's energy flow, as indicated in the inset of the right panel in Fig. 2B by the variation in the confinement direction (here the *z*-direction) of the magnitude of the Poynting vector in the direction of propagation. While the energy flow of the GSP is bound to the graphene layer, the IGP mostly propagates in the (low-loss) dielectric layer, and is thus predicted to be less sensitive to loss in graphene when nonlocal effects are not considered [38].

Considering the aforementioned, IGPs appear increasingly promising for applications operating at mid-IR frequencies, where graphene has a strong plasmonic response [5], feasible low-loss dielectrics are available, and gold can be approximated as a PEC. The mid-IR range is particularly important, as it covers the vibrational energies of many important molecules, carries the heat signature, and facilitates radiative heat transfer, prompting realization of numerous graphene-based plasmonic devices in recent years [44].

Another 2D crystal – monolayer black phosphorus – has been predicted to support surface plasmons [7-9], although they have yet to be observed experimentally. Naturally, black phosphorus is also expected to support image plasmons [43]. The plasmonic response of black phosphorus drastically differs between the two atomic lattice directions (along the armchair (AC) or the zigzag (ZZ) periodicity) due to the in-plane anisotropy of the surface conductivity function, demonstrated by the field profiles of the image plasmons in Fig. 2C. The plasmonic dispersion and the FOM of the surface and image plasmons (*d* = 5 nm) at mid-IR and THz energies are shown in Fig. 2D; in contrast to the IGP, the FOM of image modes in black phosphorus drops at higher frequencies as intraband Landau damping sets in [43].

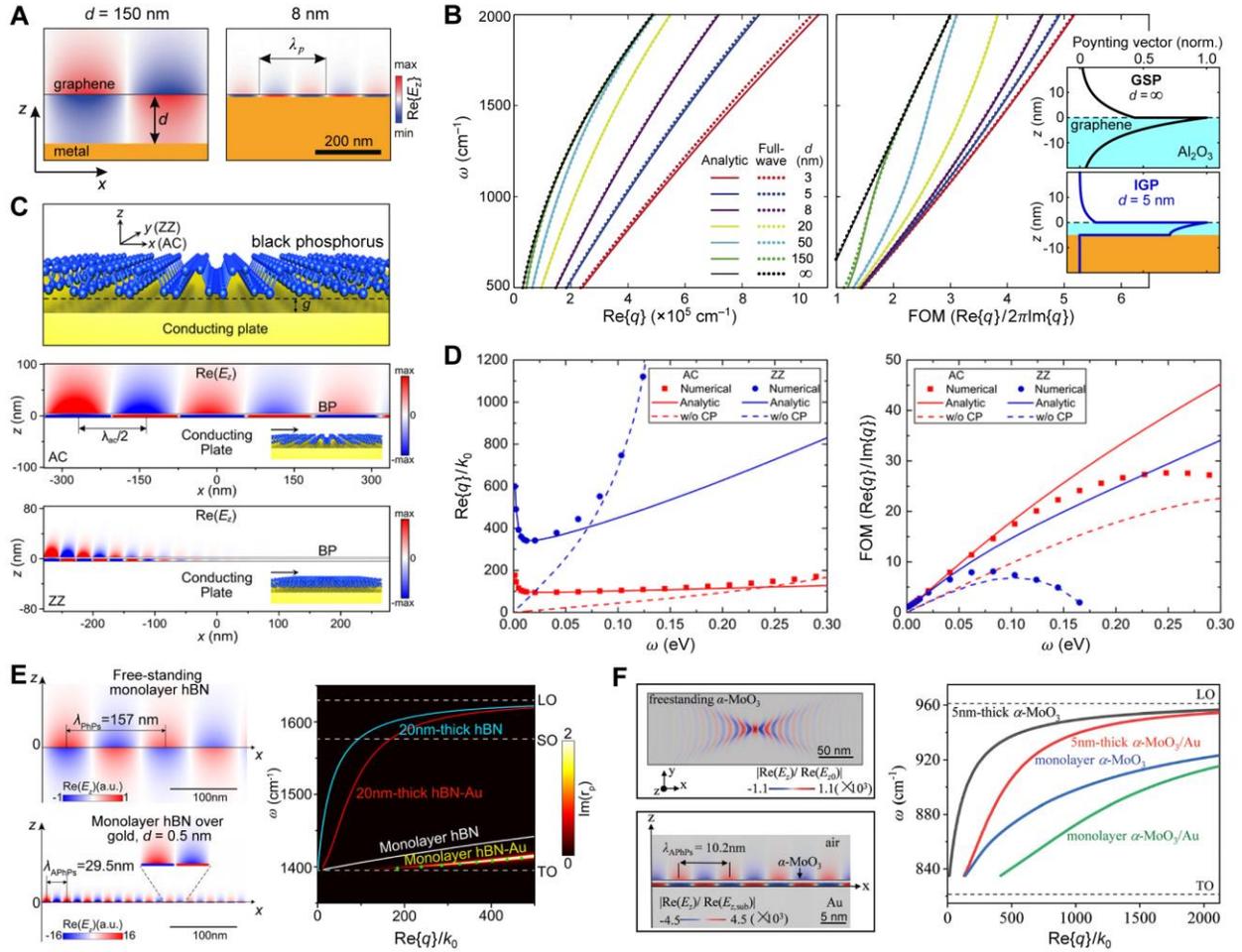

**Figure 2:** Numerical analysis of image polaritons. **A** Field profile of an image graphene plasmon (IGP): a thinner spacer leads to a more compressed IGP attaining shorter wavelengths ($E_F = 0.44$ eV). **B** Dispersion (left) and FOM (right) of mid-IR IGPs in the structure shown in **A** as a function of $d$; the inset in the right panel shows the profile of the Poynting vector in the direction of propagation for GSP and the IGP in graphene on an alumina substrate. **A,B** Adapted from [42] (inset in **B** is added by the authors). **C** Schematic illustration of the atomic structure of black phosphorus (top), and field profiles of image plasmons propagating along armchair (middle) and zig-zag (bottom) crystal axes. **D** Dispersion (left) and FOM (right) of image plasmons in black phosphorus. **C,D** Adapted with permission from [43]. Copyright 2018, American Chemical Society. **E** Field profiles (left) and dispersion (right) of the phonon-polaritons in free-standing monolayer hBN crystal, and that of the image mode emerging when hBN is placed close to a metal plate. Adapted with permission from [45]. Copyright 2020, American Chemical Society. **F** Field profiles (left) and dispersion (right) of the phonon-polaritons in the free-standing monolayer $\alpha$-MoO$_3$ crystal and the image mode in monolayer $\alpha$-MoO$_3$ above a gold plate; note that due to the in-plane anisotropy of monolayer $\alpha$-MoO$_3$, the phonon-polariton mode would have a different propagation direction in different Reststrahlen bands. Adapted with permission from [46]. Copyright 2021, Royal Society of Chemistry.

Recently, it has been predicted that a monolayer of a polar vdW dielectric can support image phonon-polaritons at frequencies within its Reststrahlen band [45, 46]. So far, such modes have been analytically studied in hBN and $\alpha$-MoO$_3$ monolayers separated from a PEC-like metal by a thin dielectric spacer. Note that such image polaritons are distinct from HIPs, which are the eigenmodes of a polaritonic slab waveguide.

A monolayer of hBN is an atomically-thin 2D crystal, and can be modelled as an isotropic 2D sheet with the effective surface conductivity $\sigma_{\text{eff}} = -i\omega h \varepsilon_0 \varepsilon_\perp$, where $h$ = 0.34 nm is the thickness of the monolayer, and $\varepsilon_\perp$ is the in-plane permittivity of hBN. Within the second Reststrahlen band ($\omega \approx$ 1370–1610 cm$^{-1}$), where Re$\{\varepsilon_\perp\} < 0$, monolayer hBN is effectively metallic, so that the field distribution and dispersion of the image modes resemble those of the IGP, as theoretically demonstrated by Yuan *et al.* [45] (Fig. 2E).

Another recently discovered vdW crystal is the $\alpha$-MoO$_3$ – a bianisotropic material with three Reststrahlen bands which supports propagating phonon-polaritons [27, 28]. Due to the in-plane anisotropy of its dielectric function, monolayer $\alpha$-MoO$_3$ supports a HPP mode that propagates in only one direction, depending on frequency. As numerically demonstrated by Lyu *at al.* [46], monolayer $\alpha$-MoO$_3$ can be effectively modelled as a film of 0.7 nm thickness; the results of full-wave simulations of the in-plane HPP in a free-standing monolayer $\alpha$-MoO$_3$ and the image modes are shown in the left panel of Fig. 2F, and their dispersion is shown in the right panel of Fig. 2F [46].

In summary, the image polaritons in monolayer vdW crystals constitute the limiting case of field compression inside a thin dielectric spacer. The analytical dispersion of image modes predicts that their effective index $n_{\text{eff}} = \text{Re}\{q\}/k_0$ can greatly exceed 100 for plasmons and phonon-polaritons at mid-IR frequencies, while their field can be strongly confined within a dielectric layer, with compression factors reaching up to ~10$^5$.

## 2.2 Image modes in multilayer van der Waals crystals

As schematically demonstrated in Fig. 1B, the second-order HPP mode of the multilayer vdW crystal is equivalent to the fundamental HIP mode in a system of a slab of half the thickness interfacing a conducting plate; compared to the relatively weakly confined fundamental HPP mode, the significantly larger momentum and FOM of the HIP provide a more appealing platform for nanophotonic applications.

The HPP dispersion relation can be found by searching for the eigenmodes of a three-layer system, in which the polaritonic material of thickness 2$t$ is sandwiched between the two dielectrics [39]. Within classical electrodynamics, the analytical form of the dispersion in a uniaxial vdW crystal can be derived under the high-momentum approximation, $q \gg k_0$, and is given by [15, 39]:

$$q = -\frac{\psi}{2t}\left[\text{atan}\left(\frac{\varepsilon_1}{\varepsilon_\perp \psi}\right) + \text{atan}\left(\frac{\varepsilon_2}{\varepsilon_\perp \psi}\right) + \pi l\right], \quad (2)$$

$$\psi = \frac{\sqrt{\varepsilon_\parallel}}{i\sqrt{\varepsilon_\perp}}$$

where $\varepsilon_1$ and $\varepsilon_2$ are the dielectric constants of the materials above and below the vdW crystal, $\varepsilon_\perp$ and $\varepsilon_\parallel$ are the in-plane and out-of-plane dielectric functions of the polaritonic material, respectively, and $l$ = 0,1,2,… indicates the mode order. Equation (2) accurately approximates the dispersion of the HIP in a slab of thickness $t$ when $l$ = 1. The dispersion relation for a non-trivial case of biaxial crystals can be found in Ref. [39].

Since phonon-polaritons in hBN exist in the two Reststrahlen bands, so do their associated image modes (Fig. 3A); figure 3B shows the analytically calculated HIP dispersion in a 100 nm-thick hBN slab on gold for the two bands. As first demonstrated by Ambrosio *et al.* [21], in the middle of the second band at ≈ 1500 cm$^{-1}$, the momentum of the first-order HIP mode is three times larger than that of the first-order HPP mode in the same hBN sample. At the same time, as first demonstrated by Menabde *et al.* [47], the FOM of the first-order image mode at the same frequency exceeds that of the fundamental HPP by ≈ 80% (Fig. 3C). Note that the FOM of the phonon-polaritons practically does not depend on hBN thickness since the propagation constant scales linearly with *t*.

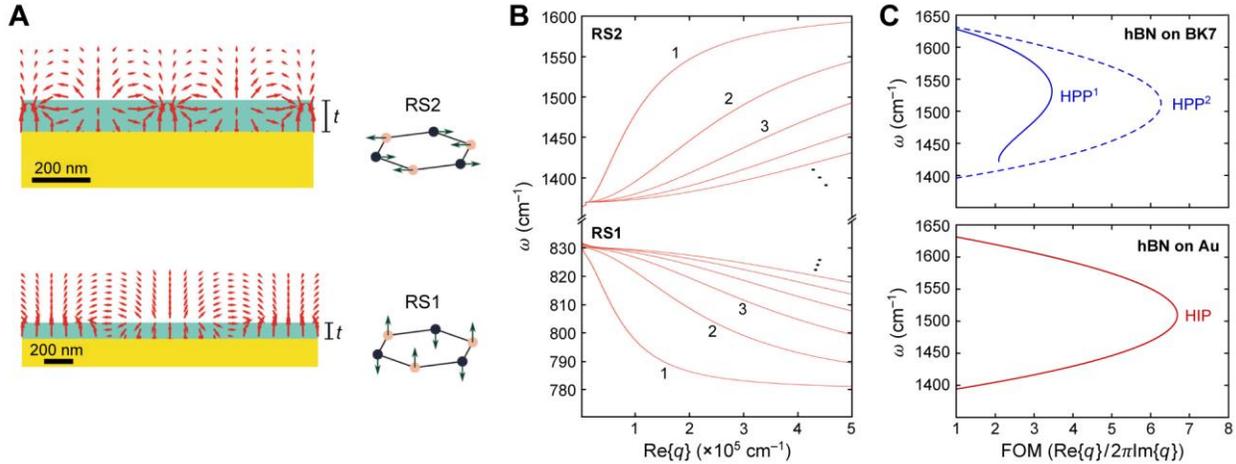

**Figure 3:** Image phonon-polaritons in hBN slab. **A** Electric field distributions of the image phonon-polaritons and corresponding hBN phonon modes excited in the two Reststrahlen bands RS1 (Re{$\varepsilon_\perp$} > 0, Re{$\varepsilon_\parallel$} < 0) and RS2 (Re{$\varepsilon_\perp$} < 0, Re{$\varepsilon_\parallel$} > 0). **B** Dispersion of image modes in the hBN slab waveguide, calculated for the two Reststrahlen bands and several mode orders. **A,B** Adapted with permission from [21]. Copyright 2018, Springer Nature. **C** Figure of merit of the fist- and second-order phonon-polaritons in hBN on glass and that of the image modes in hBN on gold, calculated for the second Reststrahlen band which is experimentally accessible by s-SNOM.

## 3. Image graphene plasmons

Graphene, the original 2D crystal, hosts highly confined and electrically tunable plasmons that have been in the spotlight within the photonics community ever since their discovery. Hence, it is only natural that the first image polariton mode was first predicted [30] and then experimentally detected [31] in graphene. The tunable and strong plasmonic response of graphene, together with the PEC-like gold conductivity at THz and mid-IR frequencies, provide a convenient platform for probing of image plasmons by near- and far-field methods. In this section, we summarize works that reveal the unique properties of the IGP, starting with near-field experiments, and then presenting far-field studies, including the demonstration of record-breaking field compression and the application of the IGP for molecular sensing.

### 3.1 Near-field probing

The scattering-type scanning near-field optical microscope (s-SNOM), which is partly based on technology from the atomic force microscope (AFM), is able to launch highly confined polaritonic modes and directly measure their near-fields [48]. The key advantage of the s-SNOM is that the coupling to the near-field is facilitated by the elastic light scattering at the AFM nano-tip, bypassing the momentum mismatch bottleneck and providing a very high spatial resolution (~10 nm) that is independent of excitation wavelength. Therefore, the s-SNOM has been extremely useful for experimental investigation of confined polaritonic modes in low-dimensional materials. At the same time, the performance of the s-SNOM is proportional to the penetration depth of the evanescent field above the sample, since the amplitude of the near-field signal *s* is directly proportional to the amplitude of the electric field: $s \propto |E_z|$ [49].

Because of the strong confinement of the IGP field inside the dielectric spacer, their first experimental detection was achieved by measuring the photo-thermoelectric (PTE) response in graphene. The PTE photocurrent in graphene is generated due to an uneven dissipation of photoexcited hot electrons at the interface between the two regions with different Fermi levels [50, 51]. Alonso-Gonzalez *et al.* [31] employed the s-SNOM to excite THz IGPs in hBN-encapsulated graphene placed onto two gate electrodes separated by a slit (top panel in Fig. 4A). In such a configuration, the electrodes provide an uneven electrostatic doping of graphene, forming a p-n junction, and simultaneously play the role of the PEC screen required for the IGP; the IGP is then launched by the AFM nano-tip and propagates to the edge of the sample, where it interferes with itself upon reflection by the edge. The near-field under the nano-tip heats the p-n junction area and generates PTE currents of different magnitude, depending on the local field intensity immediately under the tip, thus mapping the near-field interference pattern (Fig. 4A).

The detected near-field interference corresponds to a standing wave formed by the plasmon-polaritons between their source (the quasi-static nano-tip) and the reflector (typically, the sample edge) [49, 52-54]. Therefore, the periodicity of the pattern reveals the plasmon wavelength $\lambda_p$ (Fig. 4A). By imaging the near-field interference at different frequencies, the IGP dispersion can be recovered, as demonstrated in Fig. 4B for the case of a THz IGP measured by PTE mapping [31]. The photocurrent-assisted detection of the THz IGP has been also used to estimate nonlocal effects in the IGP dispersion when the spacer thickness is as small as a few nanometers [33], as will be discussed in detail in Section 5.

Recently, Menabde *et al.* [38] demonstrated near-field imaging of the mid-IR IGP in chemically-doped, large-area graphene grown by chemical vapor deposition (CVD). Similar to the photocurrent measurements, the near-field interference pattern produced by the IGP close to the graphene edge was detected. The direct optical sensing of the mid-IR IGP is made possible by the finite penetration depth of the evanescent field 25-50 nm above graphene, which can be collected by the s-SNOM (Fig. 4C). The near-field map of the graphene edge on the gold/$Al_2O_3$ substrate is shown in Fig. 4D, where the clearly visible IGP interference fringes near the edge reveal the plasmon wavelength (Fig. 4E). The dispersion of the mid-IR IGP measured in the sample with $d = 21$ nm is shown in Fig. 4F.

By contrasting the measured near-field interference profile to the results of full-wave numerical simulations, it is possible to extract the optical conductivity of graphene, and thus quantify the plasmon damping [38, 53]. The authors of Ref. [38] reported an FOM of 2.12 for an IGP excited at 1150 cm$^{-1}$ in two different samples with 18 and 8 nm-thick $Al_2O_3$ spacers, which is 1.4 times larger than the expected value for the GSP in graphene under similar conditions. At the same time, the measured IGP wavelength is 1.7 times (for $d = 18$ nm) and 2.3 times (for $d = 8$ nm) shorter than that of the GSP. These results experimentally demonstrate the predicted properties of image plasmons: simultaneously larger FOM and in-plane momentum compared to that of the GSP.

Furthermore, the near-field imaging of IGPs in a periodic array of gold nanoribbons embedded in $Al_2O_3$ revealed an excitation of the IGP Bloch state in a 1D periodic medium formed by the nanoribbons [38].

The existence of the Bloch state has been associated with the phase-matching condition between the array and the plasmonic modes propagating perpendicular to the ribbons – the IGP mode in graphene above the gold nanoribbons and the GSP mode between the nanoribbons [55]. The phase-matching condition in the array can provide an efficient far-field coupling to IGP modes even when spacer is extremely thin, as demonstrated by several studies discussed in the next two subsections.

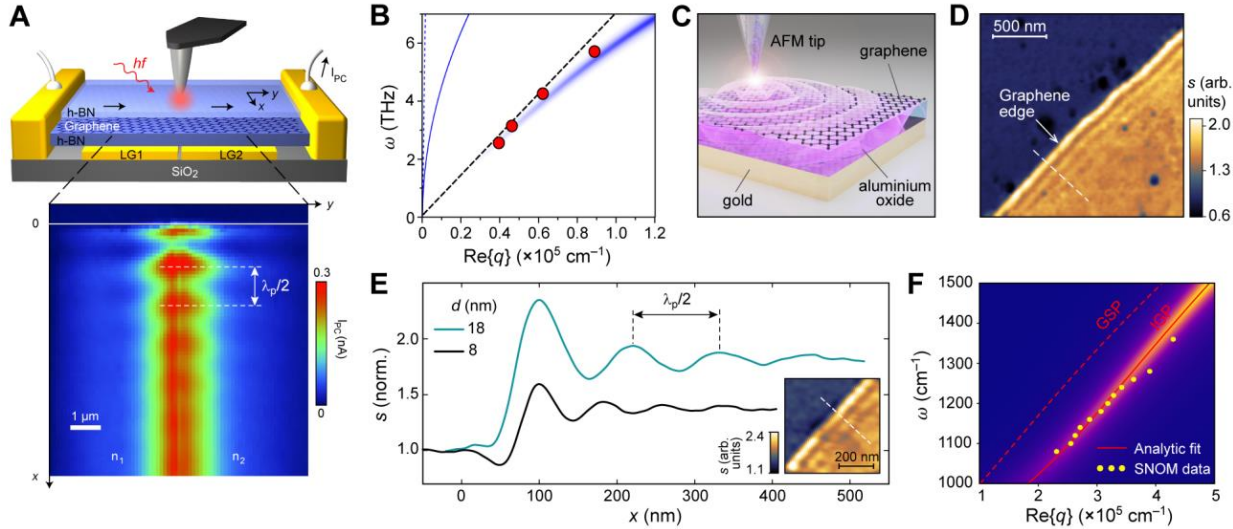

**Figure 4:** Near-field probing of image graphene plasmons. **A** PTE current-assisted near-field imaging of THz image plasmons in graphene: near-field interference modulates the PTE current generated at the p-n junction between the two graphene areas of different doping. **B** Dispersion of the THz IGP (data points) measured from the near-field interference fringes shown in **A**. The blue color plot shows the calculated dispersion of IGP in an air/hBN/graphene/hBN/AuPd/SiO$_2$ heterostructure assuming the experimental conditions, while the thin blue solid curve indicates the dispersion of GSP in free-standing graphene with the same carrier concentration. The dashed black line displays the analytical approximation of the IGP dispersion, while the dashed blue line indicates the light line in free-space, and is nearly indistinguishable from the y-axis. **A,B** Adapted with permission from [31]. Copyright 2017, Springer Nature. **C** Schematics of near-field optical probing of image graphene plasmons in the gold/alumina/graphene multilayer structure; the sharp AFM tip of the s-SNOM couples to the near-field of the plasmon-polaritons via scattering. **D** Near-field map of the graphene edge on gold/alumina substrate, spacer thickness $d = 18$ nm and $\omega_0 = 1150$ cm$^{-1}$; interference fringes of image plasmons are clearly visible, revealing the plasmon wavelength. **E** Interference fringes of image plasmons upon their reflection at graphene edge, measured in samples with $d = 18$ and 8 nm. Inset: near-field map of the graphene edge in the sample with $d = 8$ nm. **F** Measured (data points) and calculated (color map and solid red line) dispersion of an image plasmon, presented along with that of surface graphene plasmon (dashed red line); $d = 21$ nm. **C–F** Adapted with permission from [38]. Licensed under a Creative Commons Attribution.

## 3.2 Far-field probing

Near-field imaging by s-SNOM is a powerful method to investigate dispersion and loss mechanisms of propagating polaritons on nanoscopic length scales. However, strong interaction with the free-space mode is required for applications such as molecular sensing, photodetection, and active light manipulation. The large momentum mismatch between the free-space light and the mid-IR IGP can be overcome by employing arrays of metallic nanoribbons, which simultaneously act as a PEC screen and satisfy the phase-matching condition between the far-field and the plasmonic modes in the array. Upon phase matching, each nanoribbon can be considered as an IGP cavity where a Fabry-Perot resonance can be excited.

In the work by Iranzo *et al.* [32], an array of gold nanoribbons was deposited on top of a Si/SiO$_2$/graphene/dielectric structure, with the Si/SiO$_2$ substrate facilitating the electrostatic doping of graphene, as shown schematically in Fig. 5A. Most importantly, such a device configuration enabled the deposition of extremely thin dielectric layers on top of graphene: a monolayer of hBN and a 2 nm-thick layer of Al$_2$O$_3$. The far-field response of these devices was measured using Fourier transform infrared (FTIR) microscopy, as shown in Fig. 5B, where FTIR extinction spectra of a device with 256 nm-wide gold nanoribbons and 2 nm-thick Al$_2$O$_3$ spacer are presented for different Fermi levels in graphene. Since the ribbon width is larger than the IGP wavelength in such a narrow spacer, several Fabry-Perot resonances of higher orders are visible in the spectra, as noted by the order number in Fig. 5B. As the Fermi level increases, the resonance peaks exhibit a blueshift due to larger $\lambda_{IGP}$. The presence of several high-order Fabry-Perot resonances once again highlights the low normalized loss of the highly compressed image plasmons, which can oscillate up to several optical cycles before they fully dissipate.

In order to observe the first-order resonance of the ultra-confined IGP, the authors of Ref. [32] used devices with narrow gold nanoribbons. Figure 5C demonstrates the FTIR extinction spectra obtained in the device with 33 nm-wide nanoribbons and a monolayer hBN spacer with $d \approx 0.7$ nm. Variation of the graphene Fermi level reveals the blueshift of the IGP resonance as the wavelength increases along with the doping, demonstrating excitation of the IGP with the lateral field compression factor exceeding 10,000, while the effective mode index reaches 150. Upon such extreme plasmon confinement, nonlocal effects in both the graphene and the metal ultimately start to take effect. However, a complete theoretical treatment of nonlocal effects remains to be formulated; we summarize the discussion of nonlocal effects in Section 5.

Far-field excitation of IGPs in a nanoribbon array is only possible when the impinging optical field contains a polarization component aligned perpendicular to the ribbons, so that maximal coupling efficiency occurs for linearly polarized light in one direction. IGP confinement by nanoribbons is also limited to a single direction, and thus does not provide a truly localized resonance mode such as that supported by a 3D plasmonic cavity. Epstein *et al.* [56] have demonstrated that the polarization-independent excitation of localized IGP resonances is possible via the graphene plasmon magnetic resonance (GPMR) under metallic nanocubes (Fig. 5D). The local GPMR resonance can be excited by randomly depositing silver nanocubes on hBN-encapsulated graphene supported by a Si/SiO$_2$ substrate that also facilitates electrostatic doping of graphene. Figure 5E shows the FTIR extinction spectra as a function of gate voltage, indicating the presence of the IGP resonance, which is the only loss mechanism in the system which depends on doping. The IGP peak splits into two peaks due to strong hybridization with SiO$_2$ phonons (marked by the shaded area). The IGP resonance in three different samples with the nanocubes of 50, 75, and 110 nm side lengths is shown in Fig. 5F (stars), overlapping with the calculated dispersion of the uniform structure (color map). Note that the dispersion lies very close to the experimental limit given by the Fermi velocity of graphene electrons $v_F = 10^6$ m/s, corresponding to the IGP wavelength $\lambda_{IGP} = \lambda_0/300$ (dashed line in Fig. 5F). The experimentally obtained maximum mode confinement under the 50 nm cubes (lower black star) corresponds to the IGP with group velocity $\approx 1.42\times10^6$ m/s. In terms of the normalized mode volume, $V_{GPMR}/\lambda_0^3$, the single IGP cavity provides an exceptionally large mode confinement, up to $\sim 5\times10^{10}$, so that even at the mid-IR frequencies, light can be confined within a nm-scale volume.

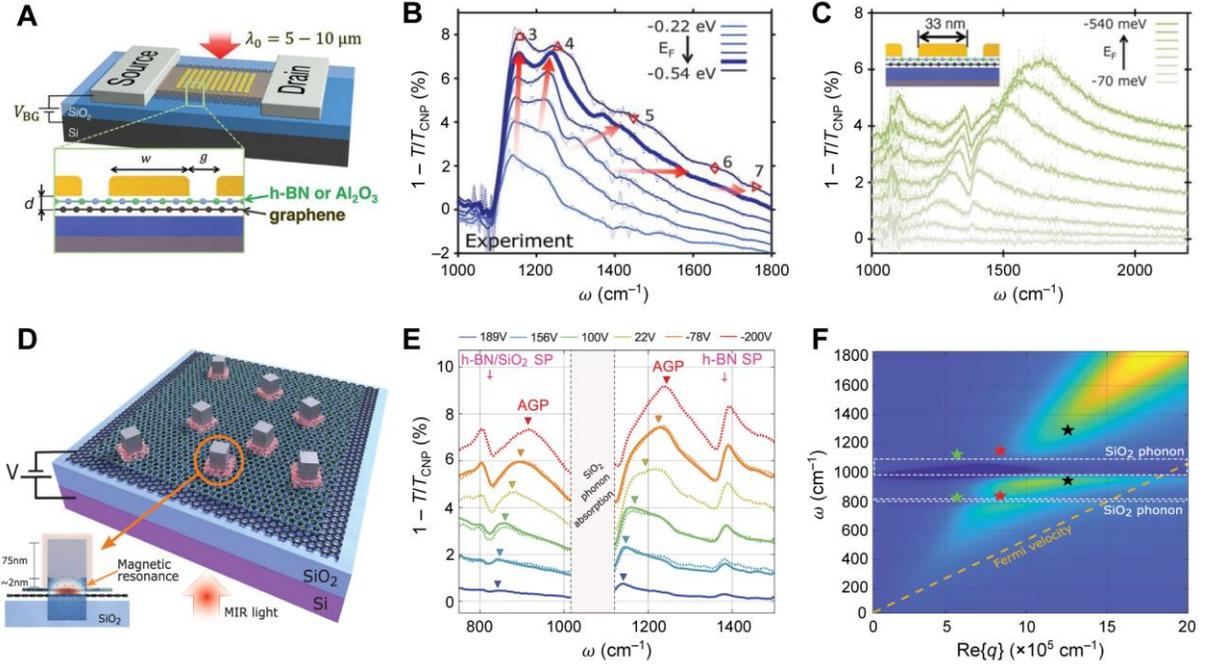

**Figure 5:** Far-field probing of image graphene plasmons. **A** Schematics of the device for the spectroscopic FTIR detection of image graphene plasmons coupled to the free-space light via the array of gold nanoribbons. **B** Extinction spectra of the device in **A** with 256 nm-wide gold nanoribbons and 2 nm-thick $Al_2O_3$ spacer at different graphene doping levels; numbers indicate signatures of higher-order Fabry-Perot resonances of image plasmons formed under the nanoribbons. **C** Extinction spectra of the device in **A** with 33 nm-wide gold nanoribbons, showing the spectral signature of only the first-order Fabry-Perot resonance. **A–C** Adapted with permission from [32]. Copyright 2018, AAAS. **D** Silver nanocubes randomly deposited on the $Si/SiO_2$/graphene/hBN structure, where the localized resonance of the image graphene plasmons is excited. **E** Extinction spectra of the device in **D** for different graphene doping levels. **F** Dispersion of the localized plasmon mode (stars) in the cavity under the silver nanocubes of 50, 75, and 110 nm size; the color map indicates the dispersion of the image mode in a uniform structure. The dashed yellow line shows the experimental limit given by the Fermi velocity of free electrons in graphene. **D–F** Adapted with permission from [56]. Copyright 2020, AAAS.

The extraordinarily large momentum of the localized IGP within a nm-thick spacer implies of the need for a phase-matching mechanism to achieve far-field coupling. However, the authors of Ref. [56] have demonstrated that the nanocube arrays possess neither periodicity, nor inter-resonator interactions. They attributed the efficient far-field coupling to the presence of a magnetic dipole resonance within the nanocubes (inset in Fig. 5D), which resembles the behavior of a rectangular patch antenna placed above the grounded conductive plate often used in the radio-frequency regime. Such an antenna supports a Fabry-Perot-like resonance, and can be described by a magnetic surface current. In the structure shown in Fig. 5D, graphene simultaneously supports the IGP and plays a role of the conductive plate. Thus, far-field illumination can directly excite the GPMR patch antenna mode in the silver nanocubes, which is associated with the excitation of IGP in optical cavity formed between the graphene and the nanocubes.

### 3.3 Far-field molecular sensing by image plasmons

As discussed in the Section 3.2, periodic arrays of metallic nanoribbons provide efficient far-field coupling to the IGP beneath the metal, where graphene is electrostatically doped via the substrate. In this configuration however, the IGP resonance leads to less than 10% modulation of the FTIR extinction spectra (Fig. 5). In order to improve the far-field coupling efficiency, Lee *et al.* [55] proposed an array-based IGP resonator coupled with the optical cavity operating in the reflection regime (Fig. 6A). The presence of the optical cavity significantly increases the absorption of the mid-IR light at resonance, theoretically reaching almost 100%, as also demonstrated in previous studies employing GSP resonances [57, 58]. The experimentally demonstrated absorption reaches as high as 94%, even for a very thin spacer of thickness $d$ < 10 nm (Fig. 6B). Furthermore, commercially available large-area CVD graphene was used, being exposed to ambient conditions as it was doped chemically. Yet again, the almost perfect absorption in the resonator with unprotected CVD graphene of moderate quality highlights the large FOM of the IGP and suggests lower sensitivity to the loss in graphene.

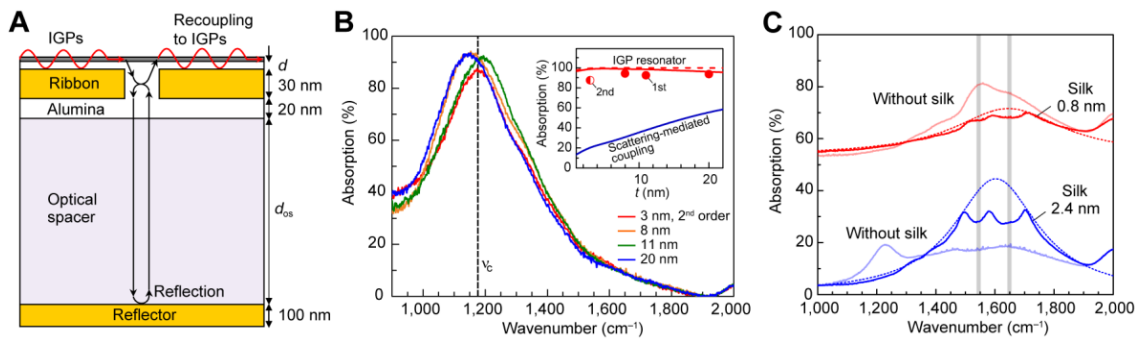

**Figure 6:** Molecular sensing with image graphene plasmons. **A** Image plasmon resonator with gold nanoribbons embedded in the alumina and the optical cavity. **B** Absorption spectra of the IGP resonator with different alumina spacer thicknesses and ribbon widths, providing phase matching between the optical cavity and the IGP resonance in the periodic structure. **C** Absorption spectra of the IGP resonator with (thick solid) and without (thick solid) a thin silk film; absorption lines of silk amides are clearly visible even for the 0.8 nm-thick film. Adapted with permission from [55]. Copyright 2019, Springer Nature.

The large surface area and high coupling efficiency of the proposed IGP resonator have been demonstrated to provide an effective platform for molecular sensing, where the ultra-compressed field of the IGP supplies the requisite strong light-matter interaction. As an example, authors of Ref. [55] demonstrated the detection of the sub-nm-thick films of silk (0.8 nm) and $SiO_2$ (0.2 nm). In particular, the IGP resonator revealed the absorption signature of silk amides, with an absorption modulation of about 3.6% for a 0.8 nm-thick silk film, and reaching as high as ≈ 13% for a 2.4 nm-thick silk layer (Fig. 6C). At the same time, the control sample with 2.4 nm of silk without the IGP resonator demonstrated an order of magnitude weaker absorption signature of less than 1%.

## 4. Hyperbolic image phonon-polaritons

While graphene requires external doping to support low-loss mid-IR IGPs, polar dielectric materials, such as hBN and $α$-$MoO_3$, support phonon-polariton modes within their Reststrahlen bands that are easily accessed in a passive configuration with no need for doping. In particular, the second Reststrahlen band of hBN (approximately at 1400–1600 cm$^{-1}$, where Re$\{ε_⊥\} < 0$ and Re$\{ε_∥\} > 0$) is spectrally located within the mid-IR band of the quantum cascade lasers and can be accessed by the s-SNOM. Although both the dispersion and the FOM of the HPPs in hBN have been extensively investigated by the near-field experiments [2, 10-26], the case of HIP has been rarely considered [20, 21, 47]. Notably, the HIP exhibits dispersion properties very similar to that of the IGP: shorter wavelength, yet longer normalized propagation length. In both cases, the behavior is mostly due to a significant reduction of the polariton group velocity while its lifetime is practically unchanged.

**4.1 Near-field probing**

When the s-SNOM nano-tip is the only metallic scatterer, the near-field interference pattern corresponds to that of a standing wave formed by the roundtrip of polaritons between the tip and the material edge [15, 16, 19, 21, 22, 24, 25], similar to the near-field probing of graphene plasmons. However, the thick edge of the hBN slab serves to scatter the illumination beam and launch polariton modes, while the material itself also contributes to the near-field signal, resulting in an overall signal containing contributions from multiple sources, as schematically shown in Fig. 7A [15, 19, 21, 22]. This, along with the generally diverging wavefront of arbitrary shapes, renders the analysis of near-field images prohibitively complicated [26, 59]. Recently, an alternative approach to the near-field probing of HPP and HIP based on large monocrystalline gold flakes [60] has been suggested, enabling accurate measurements of their complex propagation constants [47].

The momentum of HIPs in hBN has been measured in both the first and second Reststrahlen bands. Since the first band (780–830 cm$^{-1}$) lies outside of the currently available spectrum of laser sources, Ambrosio *et al.* [21] used photoinduced force microscopy (PiFM) [61-64] to image the HIP in the near field at these frequencies, while using both PiFM and s-SNOM for the second Reststrahlen band. The HIP near-field mapping by PiFM in the hBN on a gold substrate is shown in Fig. 7, where the slightly irregular profile of the interference fringes reveals the presence of two HIP modes: the one launched by the tip and the one launched by the material edge. In this case, the HIP dispersion can be extracted by correcting for the diverging wavefront of the tip-launched mode and the consequent Fourier analysis of the interference fringes with spectral filtering [21, 22]. The measured HIP momenta in the first Reststrahlen band, along with the analytical solution, are presented in Fig. 7C. Note that the hyperbolic modes have anomalous dispersion (i.e. both negative group velocity and effective refractive index).

Figure 7D shows the near-field image of the HPP and HIP measured by s-SNOM above and near the gold plate, as reported by Duan *et al.* [20]; the extracted polariton momenta and the analytically calculated dispersion are shown in Fig. 7E. Since both the HPP and the HIP are launched by the gold edge having a much larger scattering cross-section [19], and due to the weak reflection of the tip-launched modes at the gold edges [47], the contribution from the tip-launched modes to the near-field interference is relatively small. Therefore, the interference pattern is a result of the field superposition between the propagating polaritons and the quasi-uniform illumination beam of the s-SNOM, and the period of interference fringes is equivalent to the polariton wavelength [19]. Although the polariton momentum can be directly obtained from the interference fringes, extraction of the FOM remains a non-trivial problem.

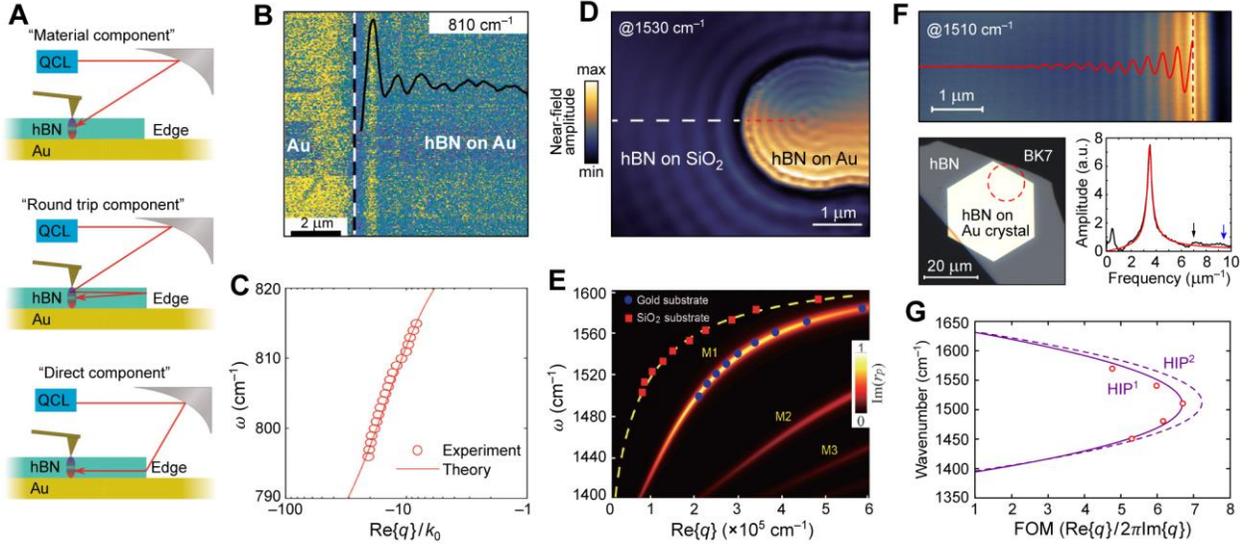

**Figure 7:** Near-field probing of image phonon-polaritons in hBN. **A** Different components of the near-field signal detected by s-SNOM when scanning the hBN edge on gold. **B** Near-field map of the hBN edge measured by PiFM in the first reststrahlen band RS1; the black solid line indicates the profile of the near-field interference fringes formed by image phonon-polaritons upon their reflection by the hBN edge. **C** Dispersion of image phonon-polaritons in a hBN slab obtained from the PiFM near-field maps in RS1. **A–C** Adapted with permission from [21]. Copyright 2018, Springer Nature. **D** Near-field map of the phonon-polaritons in hBN partially covering the gold pad, measured by s-SNOM in the RS2 band. **E** Dispersion of phonon-polaritons in hBN on $SiO_2$ and image modes in hBN on gold, obtained in the RS2 band by s-SNOM as shown in **D**. **D,E** Adapted with permission from [20]. Copyright 2017, Wiley. **F** Near-field probing of the image polaritons in hBN on gold crystal (bottom left); the straight edge of the gold crystal launches polaritons with planar wavefront (top), so that their complex propagation constant can be recovered from the Fourier spectrum of the interference fringes (bottom right). **G** Measured (data points) and calculated (solid) FOM of the image polaritons in hBN; perfect agreement between the experimental and analytical data demonstrates the absence of scattering loss due to the atomically smooth surface of the gold crystals.

Very recently, Menabde *et al.* [47] suggested to use large-area gold crystals as an ultra-low-loss substrate for mid-IR image polaritons. The straight and long (~20 μm ≫ $\lambda_p$) edges of gold crystals launch polaritons with a planar wavefront, so that the interference fringes directly correspond to the exponentially decaying field amplitude of the propagating HIP modes launched by the edge (Fig. 7F). Another advantage of this platform is that the interference fringes can be integrated across large scan areas, which significantly improves the data quality. Fourier analysis of the fringe profiles produce very clean signals that can be straightforwardly fitted to the Lorentzian spectrum of a damped oscillator (Fig. 7F), readily providing the complex propagation constant of the HIP: Re{$q$} is given by the Lorentzian peak position, while Im{$q$} is given by the full width at half maximum of the peak. The huge quality of the Fourier signal is evident from the practically absent double-frequency peak of the tip-launched HIP (black arrow in the bottom right panel of Fig. 7F) and the trace signal of the ultra-compressed second-order HIP mode corresponding to the fourth-order HPP (blue arrow in bottom right panel in Fig. 7F).

The accuracy of the method suggested in Ref. [47] is demonstrated by Fig. 7G, showing a perfect agreement between the measured and the numerically calculated FOM of the HIP, where the latter is obtained using the independently recovered hBN dielectric function. Interestingly, the FOM of the HIP exhibits a parabolic spectral dependency. At the frequency of maximal FOM, the HIP propagates 1.9 times further in terms of optical cycles than the HPP on the dielectric substrate. At the same time, the HIP has half the wavelength of the HPP. Such behavior is notably similar to the IGP, and in both cases, is mainly due to the rapid reduction in the group velocity at practically constant polariton lifetime.

## 4.2 Far-field probing

The fundamental HIP mode has significantly larger momentum compared to the fundamental HPP mode in the same polaritonic waveguide, and thus cannot be easily excited in far-field regime. As we discussed in Section 3.3, it is possible to obtain a nearly 100% far-field coupling efficiency to the ultra-compressed IGP with the aid of a metallic array and an optical cavity. Lee *et al.* [40] applied this approach to demonstrate efficient far-field coupling to HIP modes in a gold/Al$_2$O$_3$/hBN heterostructure. There, the presence of the spacer provides the condition that allows the existence of both symmetric and antisymmetric image modes (Fig. 8A), while the hBN on gold supports only the symmetric mode (as discussed in Section 4.1). The symmetric mode is mostly localized in the spacer between the hBN and gold, and hence, is sensitive to its thickness (Fig. 8B). On the other hand, the field associated with the anti-symmetric mode is mostly confined within the hBN slab and is thus weakly sensitive to the hBN-gold separation distance (Fig. 8B). Furthermore, the anti-symmetric mode has significantly larger effective index (Fig. 8C), which explains the weaker FTIR signature compared to that of the symmetric mode (Fig. 8B,D).

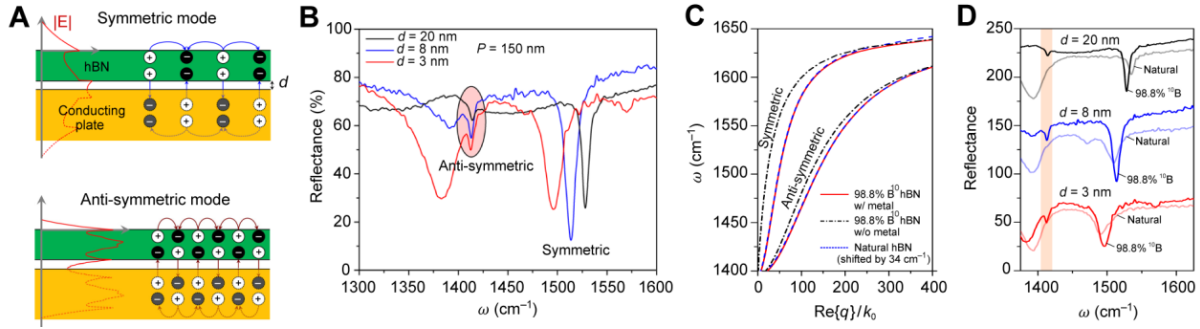

**Figure 8:** Far-field probing of the image phonon-polaritons in hBN. **A** Charge distribution of the symmetric and anti-symmetric image modes in hBN separated from gold layer by a thin alumina spacer of thickness $d$. **B** Reflectance spectra of the isotopically enriched h$^{10}$BN deposited on the resonator structure with the optical cavity shown in Fig. 6A, for different alumina thickness $d$ and the same period $P$ of the nanoribbons. **C** Calculated dispersion of the symmetric and anti-symmetric image modes in hBN of the two isotopic compositions: naturally abundant (blue) and isotopically enriched (red); the black dot-dashed line indicates the dispersion of a free-standing isotopically enriched hBN slab. **D** Reflectance spectra of the resonators with different alumina thickness, measured for natural and isotopically enriched hBN samples; the shaded resonance peak corresponds to the high-index anti-symmetric mode, present only in the isotopically enriched h$^{10}$BN due to much lower phonon damping. Adapted with permission from [40]. Licensed under a Creative Commons Attribution.

The authors of Ref. [40] analyzed the HIP resonance in isotopically enriched hBN containing 98.8% of $^{10}$B isotope, in the spirit of earlier reports that the HPP in enriched hBN possesses up to five times larger lifetime than the HPP in naturally abundant crystals with typical isotopic compositions of $^{10}$B:$^{11}$B ≈ 20%:80% [22]. Consequently, the HPP modes in isotopically enriched hBN have a few times larger FOM [22, 24]. Using the HIP resonator structure with an optical cavity and low-loss h$^{10}$BN, the authors of Ref. [40] reported the far-field detection of HIP modes with record-high effective index and quality factor: $n_{eff}$ = 132 and $Q$ = 501 for the anti-symmetric mode, and $n_{\text{eff}} = 85$ and $Q = 262$ for the symmetric mode. Notably, the far-field signature of the strongly confined anti-symmetric mode is not observed in natural hBN, but is present in the isotopically enriched samples (Fig. 8D). Note that the different intensity of the reflectance dip in Figs. 8B,D is due to the mismatch between the resonance frequencies of the optical cavity and the HIP Fabry-Perot resonance in the periodic structure when $d = 20$ and 3 nm.

## 5. Nonlocal effects revealed by image plasmons

In the previous sections, we have shown that the IGP can be effectively squeezed into atomic-scale volumes. Classical electrodynamics predicts that, when the graphene-metal separation distance $d$ is compressed to such minuscule scales, the plasmon velocity becomes proportionally smaller (Fig. 9A); this reduction is naturally accompanied by exceptionally large plasmon wavenumbers (Fig. 9B). The first indication of shortcomings in the classical description is that the plasmon wavenumbers may seemingly exceed the Fermi wavenumber of the underlying electronic system. The classical description, however, fails to consider intrinsic nonlocal effects, amounting to significant deviations between classical predictions and the proper nonlocal plasmon dispersion. These underlying nonlocal effects, possibly in both the 2D material and the supporting metal substrate, manifest in wavevector-dependent response functions (surface conductivities and dielectric functions), thus also commonly referred to as spatial dispersion. In fact, as $d$ approaches a few nanometers, classical theory predicts plasmon velocities below the Fermi velocity of electrons, or holes, in graphene, hence the plasmon dispersion diagrams situated well within the electron-hole continuum. Therefore, it is crucial to explore how nonlocal descriptions of IGPs are employed to secure adequate predictions.

### 5.1 Nonlocal and quantum graphene plasmonics

The standard tactic for dealing with nonlocality and the full wave-vector dependence in graphene has been to employ the nonlocal random-phase approximation (RPA), which provides a highly accurate interpretation of the electrodynamics affecting graphene plasmonics [6]. Though accurate, the nonlocal RPA is still just a first approximation of the full quantum mechanical response, and notably neglects electron-electron correlations. Many-body corrections to the RPA become gradually more pronounced for low doping, and were experimentally reported in the pioneering experimental study by Lundberg *et al*. [33] where a renormalization of the Fermi velocity and a compressibility correction was suggested.

The configuration employed by authors of Ref. [33] is very much like the standard IGP heterostucture (Fig. 2A), i.e., a hBN-graphene-hBN-metal heterostructure (Fig. 9C). The system allows for the detection of photocurrent by creating a confined slit in the metal layer, thereby forming a p-n junction through the application of distinct gate voltages. Furthermore, the slit launches graphene plasmons excited by a terahertz beam via the s-SNOM. In the experiment, the plasmon wavenumber was extracted from the edge-reflection fringes and the plasmon phase velocity was obtained from many scanning photocurrent maps. However, the theory employing an uncorrected RPA displays a significant discrepancy between the prediction and the experimental result. In order to address this discrepancy, many-body corrections have been employed.

The compressibility correction encapsulates an effect from electron liquid correlations producing a Pauli-Coulomb hole [33]; having said that, it should be noted that this appears to be largely negligible (Fig. 9D).

Fermi velocity renormalization, however, is crucial in graphene studies. It has been theoretically predicted and experimentally demonstrated that the Fermi velocity in graphene $v_F$ is significantly affected by electron-electron interactions, and depends logarithmically on the charge carrier density *n* in graphene [33, 65]. As a result, at extremely low carrier densities of $n < 10^{10}$ cm$^{-2}$, $v_F$ can reach $\sim 3 \times 10^6$ m/s [65], which is about three times higher than the conventional Fermi velocity $v_F \sim 1 \times 10^6$ m/s at typical $n \sim 10^{12}$ cm$^{-2}$ [6]. The results of Lundberg *et al*. [33] thus indicate that the RPA still provides a highly accurate description of graphene plasmons, if electron-electron correlations are accounted for with a Fermi velocity renormalization (Fig. 9D).

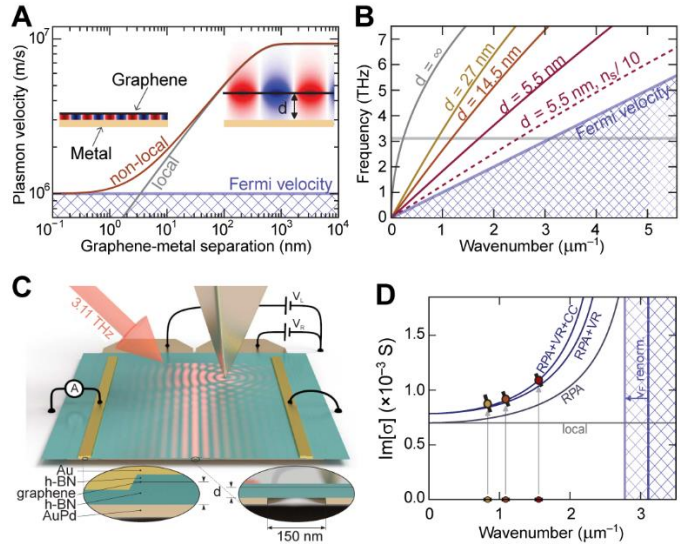

**Figure 9:** Probing nonlocal effects in graphene with IGPs. **A** Graphene plasmon velocity as a function of the separation *d* in the graphene-metal system. **B** Frequency–wavenumber dispersion of the plasmon at various *d*. **C** A metallized tip scans over a graphene sheet that has been encapsulated in hBN and placed on a split metallic film. **D** Experimentally extracted graphene conductivity $\sigma(q,\omega)$, compared with theoretical approximations for the interacting electron system in graphene: RPA, with added velocity renormalization (RPA+VR), and then with compressibility correction (RPA+VR+CC), with the local RPA appearing as a horizontal line. Adapted with permission from [33]. Copyright 2017, AAAS.

### 5.2 Probing the nonlocal response of metal interfaces

Up to this point, only nonlocal response in the graphene has been considered, while any nonlocal contributions of the adjacent materials have been neglected. The initial rationale behind this is that graphene itself dominates the electrodynamics affecting the graphene plasmons; however, the high momenta inherently carried in the IGP system dictates that, for a small *t*, the nonlocal response of the metal eventually needs to be taken into account as well.

In the Section 3.2, we discussed how Iranzo *et al*. [32] experimentally verified the necessity for considering nonlocal effects in metals by employing an array of gold nanoribbons deposited on top of a dielectric spacer/graphene/SiO$_2$/Si layered heterostructure. In order to describe the results, they developed a theoretical model accounting for nonlocal effects in both the graphene and the metal optical response. Using a local permittivity model for the metal, they suggested that nonlocality can be effectively accounted for by pragmatically increasing the spacer thickness. This phenomenon arises as a consequence of saturation of the electron density resulting in significant field penetration into the metal [66]. The vertical mode length – i.e., the ratio of energy density integrated over the out-of-plane coordinate to the maximum of the field intensity in the spacer region – is effectively limited to ∼0.3 nm by the field penetration because the energy density in the metal becomes larger than in the dielectric spacer at a certain $d$ (Fig. 10A).

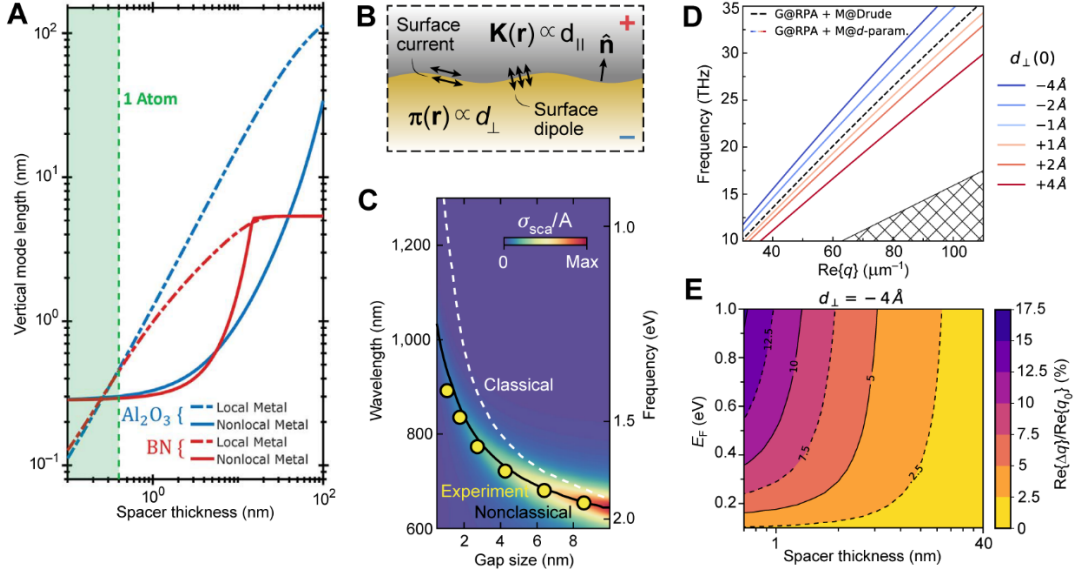

**Figure 10:** Probing nonclassical response of metals. **A** Vertical field confinement for both $Al_2O_3$ and hBN as a function of the spacer thickness for local (dash-dotted lines) and nonlocal (solid lines) metal permittivity. Adapted with permission from [32]. Copyright 2018, AAAS. **B** Nonclassical corrections can be formulated as surface polarizations: the effective surface dipole density $\pi(\mathbf{r})$ and the current density $\mathbf{K}(\mathbf{r})$. **C** Observation of large nonclassical corrections. **B,C** Adapted with permission from [67]. Copyright 2019, Springer Nature. **D** Impact of $d_\perp$ on the IGP dispersion. **E** Relative theoretically predicted shift of the IGP wavevector with and without applying the $d_\perp$-parameter correction for different graphene doping levels $E_F$ and spacer thicknesses $d$. **D,E** Adapted with permission from [36]. Licensed under a Creative Commons Attribution.

Recent works have more rigorously analyzed different nonlocal and quantum-mechanical models for plasmons in metallic films screening graphene plasmons [34, 68]; however, it may be more elucidative to rely on the formalism of surface-response functions in form of the Feibelman $d$-parameters. Recently, the classical electrodynamic boundary conditions were amended with the Feibelman $d_\perp$- and $d_\parallel$-parameters, which are mesoscopic surface-response functions with a role analogous to that of a local bulk permittivity [36, 67, 69-71]. These surface-response functions ultimately incorporate nonlocality, electronic spill-in/spill-out, and surface-enabled Landau damping to provide leading-order-accurate corrections to classical electrodynamics by driving an effective nonclassical surface polarization with an out-of-plane surface dipole density and an in-plane surface current density proportional to $d_\perp$ and $d_\parallel$, respectively (Fig. 10B). The work by Yang *et al.* [67] experimentally verified that the blueshift relative to the classical prediction, of up to 30% for small gaps, observable in gap-surface plasmon configurations can be accounted for with the Feibelman $d$-parameters (Fig. 10C). The $d$-parameter formalism has also been demonstrated to capture nonclassical effects that emerge in the optical response of crystalline noble metal surfaces, including the projected band gap emerging from atomic-layer corrugation and electronic surface states, which are predicted to influence the IGP response in heterostructures containing crystalline metals [34, 72]. While the surface-response formalism has had a revival within nanophotonics, their accurate measurements and tabulations for common interfaces remain the subject of future work [73].

Evidently, the Feibelman *d*-parameters – in particular, the $d_\perp$-parameter – should be measurable with IGPs. Gonçalves *et al.* [36] theoretically suggested doing so through a combination of numeric and perturbation theory. Specifically, they found that IGPs are spectrally redshifted for $\text{Re}\{d_\perp\} > 0$ (correlated with electronic spill-out) and blue-shifted for $\text{Re}\{d_\perp\} < 0$ (spill-in); an effect especially pronounced in the dispersion diagram of the plasmon (Fig. 10D). The relative IGP wavevector shift, $\text{Re}\{\Delta q\}/\text{Re}\{q_0\}$, with $\Delta q \equiv q_0 - q$ where $q_0$ and $q$ denote the IGP wavevector associated with $d_\perp = 0$ and $d_\perp \neq 0$, approaches 5% in the few nanometer regime and can shift a remarkable 10% for $d \gtrsim 1$ nm with modest doping (Fig. 10E). Clearly, it is important to consider the nonlocal response in metals for systems such as IGP-supporting heterostructures; and crucially IGPs themselves may provide the option for systematically measuring these effects. Finally, we mention how the concept of using the IGP to probe the electrodynamics of nearby matter even goes beyond the example of metals. As an example, Costa *et al.* [37] have suggested that the long-wave vector response of the IGP could be used to probe the electrodynamics of strongly-correlated matter, such as the Higgs mode of a superconductor.

## 6. Exciton-polaritons in semiconducting van der Waals materials on metal

The concept of image polaritons may be generalized to the situation of exciton-polaritons in semiconducting 2D materials interfacing with metal surfaces. In brief, excitons are electron-hole pairs bounded through their mutual Coulomb attraction, and they may couple strongly with light to form exciton-polaritons. Compared to plasmon-polaritons, exciton-polaritons exhibit longer coherence times, while possessing large transition dipole moments required to enable strong light-matter interactions. While excitons have been explored for decades in various 3D semiconductor compounds at low temperatures, 2D materials have stimulated renewed interest in exciton physics, largely due to the possibility for room-temperature experiments made possible by the much higher exciton binding energies available in 2D materials [74]. Typically, the pronounced exciton response in 2D materials occurs in the visible and near-infrared regimes [75, 76], where common metals host surface-plasmon polaritons that cause the metal to exhibit significant dispersive deviations from a PEC. In that respect, the problem of a 2D semiconductor on a metal offers possibilities for explorations of hybrid exciton-plasmon-polaritons [77] and strong-coupling phenomena [78, 79].

In this context, Gonçalves *et al.* [80] theoretically considered metal-oxide-semiconductor structures (with in-plane translational invariance) as illustrated in Fig. 11A to explore the hybridization of 2D excitons, such as those supported in transition metal dichalcogenides (TMDCs), with surface plasmons hosted by a semi-infinite metal substrate or a metallic thin film. By exploring additional high-index dielectric loading or tuning of the metal-film thickness, the exciton and the surface-plasmon resonances can be brought into alignment, thereby enabling plasmon-exciton interactions that enter the strong-coupling regime. Potentially, this should enable Rabi splittings beyond 100 meV in planar dielectric/TMDC/metal structures, even under ambient conditions, as shown in Fig. 11B. This is a topic undergoing significant experimental developments in recent years [81]. As another example of exciton-plasmon coupling, Shi *et al.* [82] experimentally realized metal-oxide-semiconductor structures by placing a WS$_2$/MoS$_2$ monolayer heterostructure on top of an Al$_2$O$_3$-coated single-crystalline Ag flake, demonstrating that surface plasmons may mediate the transfer of energy between the two 2D excitonic materials.

As an alternative to dielectric loading [80], plasmons may be tuned through nanostructuring of the metal and the concept of either plasmonic lattices [83] or localized plasmon resonances [84-88]. Likewise, the in-plane translational invariance of the plasmonic subsystem may be preserved, while nanostructuring the excitonic subsystem. As an example of this, Zhang *et al.* [89] experimentally demonstrated how nanostructured (multilayer) TMDC may likewise enable the desired mutual tuning of excitons and plasmons with hybridization energies exceeding 400 meV as shown in Fig. 11C and D.

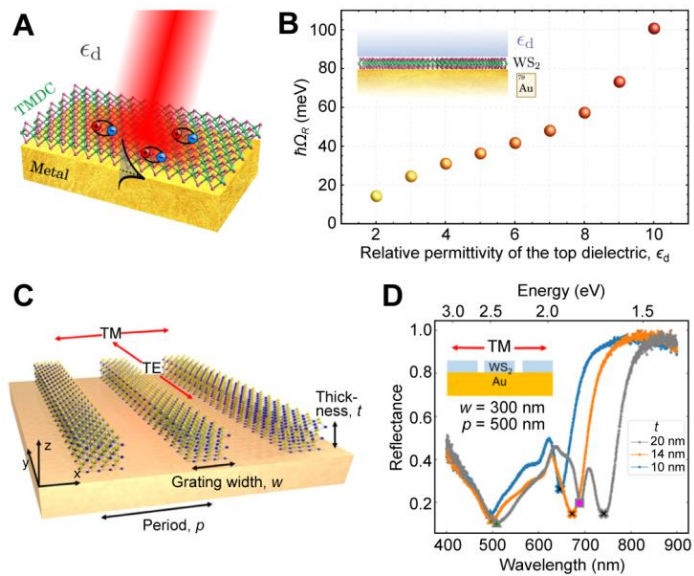

**Figure 11:** Hybridization of exciton-polaritons and metal plasmons. **A** Plasmon-exciton coupling in the dielectric/TMDC/metal system. **B** Rabi splitting energy (in meV) as a function of the dielectric constant of the top insulator in dielectric/WS$_2$/Au structures. **A,B** Adapted with permission from [80]. Copyright 2018, American Physical Society. **C** Schematic of a multi-layer WS$_2$ grating structure on a gold substrate. **D** Experimental TM reflectance spectra of the 1D WS$_2$ grating on the gold substrate with $w$ = 300 nm and $p$ = 500 nm. The peaks correspond to various resonant modes of hybrid polaritons in the grating. **C,D** Adapted with permission from [89]. Licensed under a Creative Commons Attribution.

## 7. Outlook

Here we have highlighted the key overarching properties of image polaritons in vdW crystals, as revealed by recent theoretical and experimental investigations of these hybrid nano-optical excitations. Image modes have been demonstrated to exhibit exceptionally large field confinement, with compression factors reaching up to ~10$^5$ for propagating modes, and the volume confinement factors of up to ~10$^{10}$ in cavity-localized modes. The extreme optical confinement enabled by image polaritons becomes possible due to the lack of a geometry-driven cutoff for image modes propagating within the nanometer-size gaps between the metal layer and the polaritonic material. Therefore, besides delivering a strong light-matter interaction in general, image modes can be slowed down to velocities comparable to the speed of free carriers in semiconductors and metals, thus providing access to nonlocal optical regime where classical and quantum physics coalesce.

The unparalleled field confinement in both vertical and lateral directions, along with their low loss characteristics, makes image polaritons in vdW crystals a promising platform for functional nanophotonic devices with subwavelength manipulation of light. Considering that the wavelength of image polaritons is about two orders of magnitude shorter than the wavelength of light in free space, tens of thousands image polariton resonators can be integrated in a diffraction-limited spot, offering an unprecedented degree of freedom for subwavelength light modulation. In particular, image plasmons in vdW crystals, whose properties can be dynamically tuned via electrostatic gating, can be utilized in active nanophotonic devices such as tunable metasurfaces [90-92] and ultra-compact optical waveguide switches [93-95] that have been extensively investigated using graphene plasmons.

Aside from optical sensing and high-resolution light control, the extreme field confinement offered by image polaritons can be exploited to enhance and manipulate nonlinear optical phenomena in vdW crystals [96]. At the same time, image modes exhibit significantly lower normalized propagation loss compared to "conventional" polaritons in vdW crystals on dielectric substrates, potentially enabling the development of ultra-compact nanophotonic devices where wave phenomena can be harnessed on nanometer-scale lengths. For example, image polaritons could be utilized in recently demonstrated phononic metasurfaces [97, 98] and polaritonic crystals [99-101], while providing a platform in which to probe topological phenomena in polaritons [102]. In particular, high-momentum and long-propagating image modes could be of great advantage for probing polaritonic phenomena in moiré superlattices formed by twisted double-layer vdW crystals [103-107]. Furthermore, image modes could provide a superior platform for hybrid polaritons in such heterostructures as hBN on black phosphorus [108], graphene on hBN [109, 110] and their lateral heterojunction [111], and other potentially possible 2D heterointerfaces [112].